\documentclass[pra,twocolumn,tightenlines,showpacs,nofootinbib]{revtex4}
\usepackage{bm,dcolumn,amsmath,graphicx,amsfonts,amssymb}
\usepackage{epsfig}

\begin{document}
\title{Relations between matrix elements of different
  weak interactions and interpretation of the PNC and EDM
  measurements in atoms and molecules}
 
\author{V. A. Dzuba, V. V. Flambaum, and C. Harabati}
\affiliation{School of Physics, University of New South Wales,
Sydney, NSW 2052, Australia}

\date{ \today }

\begin{abstract}
The relations between matrix elements of different P- and
T-odd weak interactions are derived.
 We demonstrate that similar
relations hold for parity nonconserving (PNC) transition amplitudes
and electron electric dipole moments (EDM) of atoms and molecules.
This allows to express P- and T-odd effects in many-electron
systems caused by different symmetry-breaking mechanisms via each
other using simple analytical formulas. We use these relations 
for the interpretation of the anapole moment measurements in cesium
and thallium and for the analysis of the relative contributions of the
scalar-pseudoscalar CP-odd weak interaction and electron EDM to the
EDM of Cs, Tl, Fr and other atoms and many polar
molecules (YbF, PbO, ThO, etc.). Model-independent limits on electron
EDM and the parameter of the scalar-pseudoscalar CP-odd interaction
are found from the analysis of the EDM measurements for Tl and YbF.
\end{abstract}
\maketitle

\section{Introduction}

The study of the parity and time invariance violation in atoms, molecules
and nuclei is a low-energy, relatively inexpensive alternative to
the high-energy search for new physics beyond the standard model (see, e.g. a
review~\cite{Ginges}). Accurate measurements of the
parity non-conservation (PNC) in atoms is one of the most promising ways of
exploring this path. It culminated in very precise measurements of the PNC
in cesium~\cite{Wood}. Interpretation of the measurements based on
accurate atomic calculations led to perfect agreement with
the standard model and put strong constrains on any new physics beyond
it~\cite{DFG02,PBD09} (see also review~\cite{Ginges} for more
detailed discussion). First unambiguous measurement of the nuclear
P-odd {\em anapole moment} was also reported in the Cs PNC
experiment~\cite{Wood}. 

At present, the study of symmetry violations in atoms and molecules goes
mostly in three major directions (see, e.g.~\cite{DF10}):
(i) the PNC measurements for a chain of isotopes; (ii) the
measurements of nuclear anapole moments; and (iii) the measurements of the
P,T-odd permanent electric dipole moments of atoms and molecules.
Interpretation of the anapole moment and EDM measurements requires
sophisticated atomic or molecular calculations. The calculations are
difficult and sometimes disagree with each other.
For example, the calculations of the thallium electric dipole moment
(EDM) by Nataraj {\em et al}~\cite{Nataraj} and by Sahoo {\em et
  al}~\cite{Sahoo08} disagree with earlier 
calculations~\cite{Liu,DF09}, calculations of the nuclear
spin-dependent PNC in cesium by Mani and Angom~\cite{Mani} disagree with
earlier calculations~\cite{Murray,Safronova,Kraft,Johnson}, etc.
The difficulties are mostly due to the strong inter-electron correlations
which need to be treated to all orders of the many-body perturbation
theory. 
Majority of publications present the effect of one particular P-
or CP-odd interaction for a specific atom or a molecule. If another
weak interaction Hamiltonian is considered, the calculations
are to be done again. 

The extraction of the nuclear anapole moment from the PNC measurements
for thallium~\cite{Vetter} also presents a problem. The analysis of
the experimental data based on atomic calculations favors a negative
value for the thallium anapole moment~\cite{Vetter,Kozlov02} while all
nuclear calculations produce positive values~\cite{DKh04}.

Since the analysis based on sophisticated atomic calculations is very
complicated it is important to have additional tools which may help to
check the calculations for consistency or even avoid the calculations.
In present paper we build such tools by studying the relations between
matrix elements for different symmetry-breaking operators. We show
that using these relations the result for a specific
operator can be easily obtained analytically if the calculations for
another operator are available. The relations can also be used to
check different calculations for consistency.
Additional benefit comes from the need to separate the effects of
different sources to the PNC or EDM of atoms or molecules in the
analysis of the experimental data. 
Recent measurement of T and P violation in YbF molecule~\cite{HindsNature} 
combined with the Tl EDM measurement~\cite{Regan} would allow
to obtain independent limits on electron EDM and CP-violating interaction.
However, the relative sign of these contributions is different in
different calculations~\cite{Pendrill,DF09}. This sign problem is
solved in the present work by calculating the ratio of the matrix
elements. Then independent limits on electron EDM and
scalar-pseudoscalar CP-odd interaction is presented.

There is an active search for the CP-violating permanent EDM of polar
molecules, such as YbF~\cite{HindsNature}, PbO~\cite{PbOe,PbO},
ThO~\cite{ThOe}, etc. Interpretation of the measurements requires
complicated molecular calculations. Most of calculations consider only
one possible source of the molecular EDM: electron EDM or
scalar-pseudoscalar CP-odd interaction. We show that the analytical
ratios of the matrix elements provide reliable link between molecular
EDMs caused by different sources. Therefore, if calculations exist for
a particular CP-odd operator, no new calculations are needed to find an
EDM caused by a different operator. 

The approach developed in this paper is easy to apply when matrix
elements of the symmetry-breaking operator between 
$s_{1/2}$ and $p_{1/2}$ single-electron states strongly dominate over
other matrix elements. This is the case  for the PNC
amplitudes between atomic $s$-states and for the EDM of atoms and
molecules caused by electron EDM or scalar-pseudoscalar CP-odd weak
interaction. We also use this approach for the thallium anapole sign
problem. 


\section{Single-electron matrix elements of weak interaction and
  many-body effects}

The PNC amplitude of an electric dipole transition between states of
the same parity $|a\rangle$ and $|b \rangle$ is equal to:
\begin{eqnarray}
   E1^{\rm PNC}_{ba}  &=&  \sum_{n} \left[
\frac{\langle b | {\bm d} | n  \rangle
      \langle n | H_{\rm PNC} | a \rangle}{E_a - E_n}\right.
\nonumber \\
      &+&
\left.\frac{\langle b | H_{\rm PNC} | n  \rangle
      \langle n | {\bm d} | a \rangle}{E_b - E_n} \right],
\label{eq:pnc}
\end{eqnarray}
where ${\bm d} = -e\sum_i {\bm r_i}$ is the electric dipole operator,
$H_{\rm PNC}$ is the operator of a P-odd CP-even weak interaction.

The EDM of an atom in state $a$ is given by
\begin{eqnarray}
   d_{a}  &=&  2\sum_{n} 
\frac{\langle a | {\bm d} | n  \rangle
      \langle n | H_{\rm CP} | a \rangle}{E_a - E_n}
\label{eq:edm}
\end{eqnarray}
where $H_{\rm CP}$ is the operator of a CP-odd weak
interaction.

States $a$, $b$ and $n$ in (\ref{eq:pnc}) and (\ref{eq:edm}) are the
many-electron atomic states. However, we can start for simplicity from
an atom with one external electron above closed shells keeping in mind
cesium atom as an example. Then in the lowest order of the
perturbation theory in residual Coulomb interaction the PNC amplitude
and the EDM of the atom are given by (\ref{eq:pnc}) and (\ref{eq:edm})
in which states $a$, $b$ and $n$ are single electron
(e.g. Hartree-Fock) states. It is easy to see that all single-electron
matrix elements of weak interaction are proportional to each
other. Indeed, only short distances contribute to the value of the
matrix elements (see appendix for analytical estimations). These
distances are $r  < r_N$ for the PNC interactions and for the
scalar-pseudoscalar CP-odd interaction ($R_N$ is nuclear radius) and
$r \le a_0/Z$ for the electron EDM operator ($a_0$ is Bohr radius). On
these distances the energy of the single-electron state can be
neglected compared to the nuclear potential $-Ze^2/r$ and Dirac
equations for all single-electron states become identical. The only
difference comes from the normalization of the states. Suppose we have two
different operators of, say, $CP$-odd weak interaction $H_{\rm CP}$
and $H^{\prime}_{\rm CP}$. If we establish proportionality between
matrix elements for a particular pair $i,j$ of single electron states
\begin{equation}
\langle i|H_{\rm CP}|j\rangle = R\langle i|H^{\prime}_{\rm
  CP}|j\rangle,
\label{eq:pro}
\end{equation}
then due to the proportionality of the wave functions on short distances,
the same proportionality, with the same value of $R$ would hold 
for any pair of single-electron states and for the total EDM of atom
(\ref{eq:edm}). Furthermore, if the $s_{1/2}-p_{1/2}$ matrix elements
of the weak interaction strongly dominate over other matrix elements
then the proportionality is not affected by the many-body
effects. Indeed, if other weak matrix elements are neglected then any
many-body expression is a sum of terms with one 
$s_{1/2}-p_{1/2}$ weak matrix element in each term. Since all of them
are proportional with the same proportionality coefficient $R$ the
proportionality would hold for the sum as well.

Table \ref{t:r} shows the ratios of the $\langle 6s_{1/2} |H_{\rm W}|
6p_{1/2} \rangle$ matrix elements for cesium atom for different
P-odd and CP-odd operators of weak interaction $H_{\rm W}$ in the
relativistic Hartree-Fock (RHF) approximation and
with dominating many-body effects included. The ratios are compared
with the analytical result presented in the appendix. The many-body
effects include core polarization (CP) and Brueckner-type
correlations. Core polarization can be understood as the change of the
self-consistent core potential due to the effect of an external field
(weak interaction in our case). Its inclusion above the Hartree-Fock
level is often called the {\em random-phase approximation}
(RPA). Brueckner-type correlations are the correlations which can be
reduced to a redefinition of the single-electron orbitals, replacing
the Hartree-Fock ones by the Brueckner orbitals
(BO)~\cite{Dzuba87}. In both cases of the P-odd and CP-odd weak
interactions the ratios of the matrix elements are stable within the
1-2\% accuracy while the matrix elements change by up to two times.

\begin{table}
\caption{The ratio of the matrix elements of the 
spin-dependent to spin-independent P-odd weak interactions ($R_1$) and
scalar-pseudoscalar CP-odd interaction to electron EDM ($R_2$) for the
$6s,6p_{1/2}$ states of cesium with and without the inclusion of
dominating many-body effects. The ratios are stable while the matrix
elements change by up to two times. Units:
$\varkappa/(-Q_W)$ for $R_1$ and $10^{-15}C^{\rm PS}/d_e$ a.u. for $R_2$.} 
\label{t:r}
\begin{ruledtabular}
\begin{tabular}{llcc}
\multicolumn{2}{c}{Approximation} & 
$R_1$ & $R_2$ \\ 
\hline
RHF & $\langle \psi^{\rm HF}_{6s}|H_W|\psi^{\rm HF}_{6p}\rangle $
& 4.78 & 8.96 \\
RPA & $\langle \psi^{\rm HF}_{6s}|H_W+\delta V_{\rm
  core}|\psi^{\rm HF}_{6p}\rangle $ & 4.88 & 8.94 \\ 
BO\tablenotemark[1] & $\langle \psi^{\rm BO}_{6s}|H_W|\psi^{\rm BO}_{6p}\rangle $
& 4.78 & 9.03 \\ 
BO+CP\tablenotemark[2] & $\langle \psi^{\rm BO}_{6s}|H_W+\delta V_{\rm
  core}|\psi^{\rm BO}_{6p}\rangle $ & 4.85 & 9.01 \\ 
\multicolumn{2}{l}{Analytical,
  eq.~(\ref{eq:r1},\ref{eq:r1-sp},\ref{eq:eedm})} & 4.84 & 9.01\\  
\end{tabular}
\tablenotetext[1]{Brueckner orbitals}
\tablenotetext[2]{Brueckner orbitals and core polarization}
\end{ruledtabular}
\end{table}

\section{Parity non-conservation}

\label{pnc}
Hamiltonian describing parity-nonconserving electron-nuclear
interaction can be written as a sum of spin-independent (SI) and
spin-dependent (SD) parts (we use atomic units: $\hbar = |e| = m_e = 1$):
\begin{eqnarray}
     H_{\rm PNC} &=& H_{\rm SI} + H_{\rm SD} \nonumber \\
      &=& \frac{G_F}{\sqrt{2}}                             
     \Bigl(-\frac{Q_W}{2} \gamma_5 + \frac{\varkappa}{I}
     {\bm \alpha} {\bm I} \Bigr) \rho({\bm r}),
\label{e1}
\end{eqnarray}
where  $G_F \approx 2.2225 \times 10^{-14}$ a.u. is the Fermi constant of
the weak interaction, $Q_W$ is the nuclear weak charge,
$\bm\alpha=\left(
\begin{array}
[c]{cc}%
0 & \bm\sigma\\
\bm\sigma & 0
\end{array}
\right)$ and $\gamma_5=\left(
\begin{array}
[c]{rr}%
0 & -I \\
-I & 0
\end{array}
\right)$  are Dirac matrices, $\bm I$ is the
nuclear spin, and $\rho({\bf r})$ is the nuclear density normalized to 1.
The strength of the spin-dependent PNC interaction is proportional to
the dimensionless constant $\varkappa$ which is to be found from the
measurements. There are three major contributions to
$\varkappa$ arising from (i) electromagnetic interaction of atomic
electrons with nuclear {\em anapole moment}~\cite{FKh80}, (ii)
electron-nucleus spin-dependent weak interaction~\cite{Novikov} , and
(iii) combined effect of the spin-independent weak interaction and the
magnetic hyperfine interaction~\cite{FKh85}
(see, also review~\cite{Ginges}). In this work we do not distinguish
between different contributions to $\varkappa$ and present the results
in terms of total $\varkappa$ which is the sum of all possible
contributions. 

Within the standard model
the weak nuclear charge $Q_W$ is given by~\cite{PDG}
\begin{equation}
Q_W \approx -0.9877N + 0.0716Z.
\label{eq:qw}
\end{equation}
Here $N$ is the number of neutrons, $Z$ is the number of protons.

The PNC amplitude of an electric dipole transition between states of
the same parity $|i\rangle \equiv |J_i F_i M_i \rangle$ and $|f
\rangle \equiv |J_f F_f M_f \rangle$ is equal to:
\begin{eqnarray}
   E1^{\rm PNC}_{fi}  &=&  \sum_{n} \left[
\frac{\langle f | {\bm d} | n  \rangle
      \langle n | H_{\rm PNC} | i \rangle}{E_i - E_n}\right.
\nonumber \\
      &+&
\left.\frac{\langle f | H_{\rm PNC} | n  \rangle
      \langle n | {\bm d} | i \rangle}{E_f - E_n} \right],
\label{eq:e2}
\end{eqnarray}
where ${\bm d} = -e\sum_i {\bm r_i}$ is the electric dipole operator,
   and ${\bm F} = {\bm I}
+ {\bm J}$ is the total angular momentum. 

Applying the Wigner-Eckart theorem we can express the amplitudes via
reduced matrix elements
\begin{eqnarray}
  E1^{\rm PNC}_{fi} &=&
      (-1)^{F_f-M_f} \left( \begin{array}{ccc}
                           F_f & 1 & F_i  \\
                          -M_f & q & M_i   \\
                           \end{array} \right) \nonumber \\
   &\times& \langle J_f F_f || d_{\rm PNC} || J_i F_i \rangle .
\end{eqnarray}
Detailed expressions for the reduced matrix elements of the SI and
SD PNC amplitudes can be found e.g. in Refs.~\cite{Porsev01} and
\cite{JSS03}. For the SI amplitude we have
\begin{eqnarray}
&&\langle J_f,F_f || d_{\rm SI} || J_i,F_i \rangle =
(-1)^{I+F_i+J_f+1}\nonumber \\ 
&& \times \sqrt{(2F_f+1)(2F_i+1)} 
\left\{ \begin{array}{ccc} J_i & J_f & 1 \\
                          F_f & F_i & I \\ 
                    \end{array} \right\}  \label{eq:si0}\\
&&  \times \sum_{n} \left[
\frac{\langle J_f || {\bm d} || n,J_n  \rangle
      \langle n,J_n | H_{\rm SI} | J_i \rangle}{E_i - E_n}\right.  \nonumber \\
&& + \left.\frac{\langle J_f | H_{\rm SI} | n,J_n  \rangle
      \langle n,J_n || {\bm d} || J_i \rangle}{E_f - E_n} \right]. \nonumber 
\end{eqnarray}
It is convenient to present the amplitude in a compact form
\begin{equation}
\langle J_f,F_f || d_{\rm SI} || J_i,F_i \rangle = C(S_1+S_2),
\label{eq:c1}
\end{equation}
where $C \equiv C(F_f,J_f,F_i,J_i)$ is the angular coefficient and
sums $S_1$ and $S_2$ do not depend on $F_f$ and $F_i$:
\begin{eqnarray}
&&S_1 = \sum_{n} \frac{\langle J_f || {\bm d} || n,J_n  \rangle
   \langle n,J_n | H_{\rm SI} | J_i \rangle}{E_i - E_n},  \nonumber \\
&& S_2= \sum_n \frac{\langle J_f | H_{\rm SI} | n,J_n  \rangle
      \langle n,J_n || {\bm d} || J_i \rangle}{E_f - E_n}. \nonumber 
\end{eqnarray}

For the SD PNC amplitude we have
\begin{eqnarray}
    && \langle J_f,F_f || d_{\rm SD} || J_i,F_i \rangle =
   \nonumber \\
     &&\sqrt{(I+1)(2I+1)(2F_i+1)(2F_f+1)/I}  \nonumber \\
    &&\times
     \sum_{n} \left[ (-1)^{J_f - J_i}
     \left\{ \begin{array}{ccc}
     J_n  &  J_i  &   1    \\
      I   &   I   &  F_i   \\                                  
     \end{array} \right\}
     \left\{ \begin{array}{ccc}
      J_n  &  J_f  &  1   \\
      F_f  &  F_i  &  I   \\
     \end{array} \right\} \right. \nonumber \\
  &&\times \frac{ \langle J_f || {\bm d} || n, J_n \rangle
     \langle n, J_n || {\bm b} || J_i \rangle }{E_n -
     E_i} \label{eq:dsd}  \\
  &&+
     (-1)^{F_f - F_i}
     \left\{ \begin{array}{ccc}
     J_n  &  J_f  &   1    \\
      I   &   I   &  F_f   \\
     \end{array} \right\}
     \left\{ \begin{array}{ccc}
     J_n  &  J_i  &  1   \\
     F_i  &  F_f  &  I   \\
     \end{array} \right\} \nonumber \\
 &&\times
     \left. \frac{\langle J_f || {\bm b} ||n,J_n \rangle
            \langle n,J_n || {\bm d} ||J_i \rangle}{E_n - E_f}  \right],
\nonumber
\end{eqnarray}
where ${\bm b}$ is the electron part of the SD weak
interaction
\begin{equation}
  {\bm b} = \frac{G_F}{\sqrt{2}}{\bm \alpha} \rho({\bm
    r})\varkappa. 
\label{eq:hsdp}
\end{equation}

Like in the spin-independent PNC amplitude (\ref{eq:si0}),
it is convenient to present the SD amplitude in a compact form
\begin{equation}
  \langle J_f,F_f || d_{\rm SD} || J_i,F_i \rangle = \sum_{i=1}^4 c_iS_i^{\prime}.
\label{eq:ssss}
\end{equation}
Here $c_i\equiv c(F_f,J_n,F_i)$ ($i=1,2,3,4$) are angular coefficients
which can be extracted from (\ref{eq:dsd}). Sums $S^{\prime}_i$ do not
depend on $F_f$ and $F_i$:
\begin{eqnarray}
  &&S^{\prime}_1 =  \sum_n \frac{ \langle J_f || {\bm d} || J_n \rangle
     \langle J_n || {\bm b} || J_i \rangle }{E_{n} -
     E_{i}} ,  \nonumber \\
  &&S^{\prime}_2 =  \sum_n \frac{ \langle J_f || {\bm d} || J^{\prime}_n \rangle
     \langle J^{\prime}_n || {\bm b} || J_i \rangle }{E_{n} -
     E_{i}} , \nonumber  \\
   &&S^{\prime}_3 =  \sum_n \frac{\langle J_f || {\bm b}
     ||J_n \rangle 
            \langle J_n || {\bm d} ||J_i \rangle}{E_{n} -
            E_{f}}, \nonumber \\ 
   &&S^{\prime}_4 =  \sum_n \frac{\langle J_f || {\bm b}
     || J^{\prime}_n \rangle J^{\prime}_n  || {\bm d} ||J_i
     \rangle}{E_{n} - E_{f}}. \nonumber
\end{eqnarray}
Equation (\ref{eq:ssss}) has more terms than (\ref{eq:c1}) due to the
electron vector nature of the nuclear-spin-dependent operator.

The total PNC amplitude can be presented in a form convenient for
extraction of the values of $\varkappa$ from the PNC measurements (see,
e.g. \cite{DF11a,DF11b}) 
\begin{equation}
E_{\rm PNC} = E(1+R).
\label{eq:pnctotal}
\end{equation}
Here $E$ is the SI PNC amplitude given by (\ref{eq:si0}) and $R$ is
the ratio of the SD to SI PNC amplitudes. In this work we are mostly
interested in the values of $R$. Using compact expressions
(\ref{eq:c1}) and (\ref{eq:ssss}) one can write
\begin{equation}
R = \frac{c_1S_1^{\prime} + c_2S_2^{\prime} + c_3S_3^{\prime} +
  c_4S_4^{\prime}}{C(S_1+S_2)}.
\label{eq:R}
\end{equation}
According to the discussion of the previous section,
ratios of the SD and SI weak matrix elements do not depend on the
principal quantum number $n$. Therefore, using the ratios
\begin{eqnarray}
 && r_1 = \langle J_n || {\bm b}|| J_i \rangle/
 \langle J_n | H_{\rm SI}| J_i \rangle, \nonumber \\  
 && r_2 = \langle J^{\prime}_n || {\bm b}|| J_i \rangle/
 \langle J_n | H_{\rm SI}| J_i \rangle, \nonumber \\  
 && r_3 = \langle J_f || {\bm b}|| J_n \rangle/
 \langle J_f | H_{\rm SI}| J_n \rangle, \label{eq:r1234} \\  
 && r_4 = \langle J_f || {\bm b}|| J^{\prime}_n \rangle/
 \langle J_f | H_{\rm SI}| J_n \rangle, \nonumber
\end{eqnarray}
we can get rid of the sums $S_1^{\prime}$, $S_2^{\prime}$,
$S_3^{\prime}$, $S_4^{\prime}$, involving the SD matrix elements and replace
them with the SI sums $S_1,S_2$:
\begin{eqnarray}
R &=& \frac{(c_1r_1 + c_2r_2)S_1 + (c_3r_3 + c_4r_4)S_2}{C(S_1+S_2)} 
\nonumber \\
& =& \frac{(c_1r_1 + c_2r_2)S_1/S_2 + (c_3r_3 + c_4r_4)}{C(S_1/S_2+1)}.
\label{eq:R1}
\end{eqnarray}
The only parameter in (\ref{eq:R}) which comes from
numerical calculations is the ratio $S_1/S_2$ of two different
contributions to the SI PNC amplitude (see
Eqs.(\ref{eq:si0},\ref{eq:c1})). All other parameters are given 
by analytical expressions. The $c_1,c_2,c_3,c_4$ and $C$ parameters
are just angular coefficients. The ratios $r_1,r_2,r_3,r_4$ can also
be approximated by analytical expressions which will be discussed below.

The expression (\ref{eq:R}) can be further simplified in an important
case of an
$ns - n's$ transition (e.g. the $6s$-$7s$ transition in Cs). On a
few per cent level of accuracy the $s-p_{3/2}$ matrix elements of the
SD PNC interaction can be neglected \cite{DF11b}. This means that
$r_2=r_4=0$. We also have $r_1=r_3=r$ and (\ref{eq:R1}) is reduced to
\begin{equation}
R = r\frac{c_1S_1/S_2 + c_3}{C(S_1/S_2+1)}.
\label{eq:Rs}
\end{equation}
Substituting $r$ from (\ref{eq:r1}) leads to
\begin{equation}
R = 4.90(1-0.073Z^2\alpha^2)\frac{c_1S_1/S_2 +
  c_3}{C(S_1/S_2+1)}\frac{\varkappa}{(-Q_W)}. 
\label{eq:Rs}
\end{equation}
$Q_W$ is the weak nuclear charge. As in (\ref{eq:R}) the only parameter
which comes from numerical calculations is the ratio $S_1/S_2$. On the
other hand, knowing the value of $R$ for at least 
two hfs components of the PNC transition is sufficient for extraction
of $\varkappa$ from the measurements. 

\subsection{PNC in cesium}

Experimental values for the two different hfs components of the parity
non-conserving $6s-7s$ transition in cesium are \cite{Wood}
\begin{eqnarray}
 E_{\rm PNC}(6s_{F=4} - 7s_{F=3}) = 1.6349(80) \ {\rm mV/cm}, \nonumber \\
 E_{\rm PNC}(6s_{F=3} - 7s_{F=4}) = 1.5576(77) \ {\rm
  mV/cm}. \label{eq:csexp}
\end{eqnarray}
To extract $\varkappa$ we use (\ref{eq:pnctotal}) and calculate $R$
for these two transitions using (\ref{eq:Rs}). To do so we note that
$J_f=J_i=J_n=1/2$, $Q_W$=-73.19 (see Eq.(\ref{eq:qw})) and take
$S_1/S_2=-0.3459$ from Ref.~\cite{Porsev}. This leads to the system of
equations 
\begin{eqnarray}
&& E(1+0.06739\varkappa) = 1.6349(80), \nonumber \\
&& E(1-0.05937\varkappa) = 1.5576(77). \label{eq:system}
\end{eqnarray}
The solution for $\varkappa$ is $\varkappa$=0.382(56). This result is in good
agreement with the values $\varkappa$=0.393(56) from Ref.~\cite{Murray} and
$\varkappa$=0.375(56) from
Ref.~\cite{Safronova}\footnote{Refs.\cite{Murray} and \cite{Safronova}
use different definition of $\varkappa$. The conversion factors are
$(I+1/2)/(I+1)$ for Ref.~\cite{Murray} and $I$ for
Ref.~\cite{Safronova}. $I$ is nuclear spin, $I=7/2$ for $^{133}$Cs.}. 
Accurate calculations similar to what is reported in our previous work
for Ba$^+$, Yb$^+$ and Ra$+$~\cite{DF11b} lead to the value
$\varkappa=0.376$ which 
is in perfect agreement with the all-order calculations of
Ref.~\cite{Safronova}. This is an illustration of the accuracy of the
analysis based on the ratio of the matrix elements. The value
$\varkappa$=0.382 coming from this analysis differs by less than 2\% from
the value $\varkappa$=0.376 coming from the accurate calculations. This
difference is due to two simplifications: (a) neglecting the
$s-p_{3/2}$ matrix elements of the SD weak interaction, and (b)
assuming that the ratio of the matrix elements is the same for all
single-electron states.

Recent relativistic coupled-cluster calculations
of the nuclear spin-dependent PNC in Cs~\cite{Mani} report the values
of the SD PNC matrix elements which are about 30\% smaller than those
of the all-order calculations of Ref.~\cite{Safronova}. This is in
disagreement not only with this work but with all earlier calculations
of the SD PNC in cesium~\cite{Murray,Safronova,Kraft,Johnson}. Given the
proportionality of the matrix elements of the SI and SD weak
interactions discussed above, the results of Ref.~\cite{Mani} are also
in disagreement with all most accurate calculations of the SI PNC in
cesium (see, e.g.~\cite{DFG02,PBD09,Porsev}). The latter calculations have accuracy
better than 0.5\% and are used to test the Standard Model in Cs PNC experiment ~\cite{Wood}
where the accuracy is 0.35 \%.   
 
\subsection{Anapole moment of thallium}

The value of the nuclear anapole moment of thallium, extracted from
the measurements of the PNC in the $6p_{1/2} - 6p_{3/2}$
transition ~\cite{Vetter}  is in disagreement with the results of
nuclear calculations (see, e.g. ref.~\cite{DKh04}). The analysis of
the experimental data based on simple single-electron approximations
gives the value $\varkappa_a=-0.22 \pm 0.30$~\cite{Vetter}. The analysis
based on sophisticated many-body calculations gives very close value of
$\varkappa_a=-0.26 \pm 0.27$~\cite{Kozlov02}. On the other hand, the
``best value'' obtained in nuclear calculations is
$\varkappa_a=0.24$~\cite{DKh04}.  To extract the sign of $\varkappa_a$ 
from the experiment  one needs the relative sign of SI and SD amplitudes. 
Here we show that simple analysis
with the use of the analytical ratio of the matrix elements of the
weak interactions supports the findings of Refs.\cite{Vetter,Kozlov02}
leaving the problem of sign disagreement unsolved.

If we keep only $s_{1/2} - p_{1/2}$ matrix elements of the weak
interaction for both SI and SD interactions, then the general
expression (\ref{eq:R1}) can be reduced to
\begin{equation}
  R(F,F^{\prime}) = r\frac{c(F,F^{\prime})}{C(F,F^{\prime})},
\label{eq:RFF}
\end{equation}
where $r$ is one of the ratios $r_i$ (\ref{eq:r1234}) which corresponds
to the $s_{1/2} - p_{1/2}$ weak matrix elements, $c(F,F^{\prime})$ and
$C(F,F^{\prime})$ are corresponding angular coefficients.
Comparing (\ref{eq:pnctotal}) to the parameterization used
in~\cite{Kozlov02} ($\mathcal{R} \equiv {\rm Im}(E1_{\rm PNC}/M1)$) 
\begin{equation}
\mathcal{R}(F,F^{\prime}) = C(Z)\left[Q_W - 6\varkappa
  \xi(F,F^{\prime})\right],
\label{eq:Cxi}
\end{equation}
and substituting (\ref{eq:r1}) into (\ref{eq:RFF}) we get for the
parameters $\xi$
\begin{equation}
\xi(F,F^{\prime}) = -0.817\left(1 -
  0.073Z^2\alpha^2\right)\frac{c(F,F^{\prime})}{C(F,F^{\prime})}.
\label{eq:xi}
\end{equation}
Corresponding values of $\xi$ are compared in Table~\ref{t:xi} with
the results of the many-body calculations of~\cite{Kozlov02}. We see
that both calculations  give very close results leaving no room for a sign
error.

\begin{table} 
\caption{Parameters $\xi(F,F^{\prime})$ of the nuclear spin-dependent
  PNC amplitude in the $6p_{1/2}(F) - 6p_{3/2}(F^{\prime})$
  transitions in thallium.}
\label{t:xi}
\begin{ruledtabular}
\begin{tabular}{cccc}
$F$ & $F^{\prime}$ & this work & Ref.~\cite{Kozlov02} \\
\hline
0 & 1 & 0.947 & 1.10 \\
1 & 1 & -0.325 & -0.462 \\
1 & 2 & -0.325 & -0.348 \\
\end{tabular}
\end{ruledtabular}
\end{table}


\section{EDM of  atoms and Molecules}

\begin{table}
\caption{The ratio of the $s_{1/2}-p_{1/2}$ matrix elements of the
  electron EDM operator (\ref{eq:eedm}) to that of the scalar-pseudoscalar
  CP-odd operator (\ref{eq:r1-sp}). Numerical results for EDM of Cs, Tl and YbF
  are also given for comparison. Units: $d_e/(C^{\rm SP} 10^{-18} e \
  {\rm cm})$. (For other isotopes $R^{\prime} = A^{\prime}R/A$)}
\label{t:rcp}
\begin{ruledtabular}
\begin{tabular}{clrr}
$Z$ & Atom & \multicolumn{2}{c}{Ratios} \\
&&\multicolumn{1}{c}{Analytical}
&\multicolumn{1}{c}{Numerical} \\
\hline
37 & $^{85}$Rb      & 228 & \\
55 & $^{133}$Cs     & 158 & 163\tablenotemark[1] \\
56 & $^{138}$Ba$^+$ & 152 & \\
70 & $^{173}$Yb$^+$ & 114 & 115\tablenotemark[2] \\
81 & $^{205}$Tl     &  89 & 83\tablenotemark[3] \\
82 & $^{208}$Pb     &  88 &  \\
87 & $^{211}$Fr     &  83 & \\
90 & $^{232}$Th     &  75 & \\
\end{tabular}
\tablenotetext[1]{Cs atom, Ref.~\cite{DF09}.}
\tablenotetext[2]{YbF molecule, Ref.~\cite{Quiney}.}
\tablenotetext[3]{Tl atom, Ref.~\cite{DF09}.}
\end{ruledtabular}
\end{table}

Table \ref{t:rcp} shows the ratio of the $s_{1/2}-p_{1/2}$ matrix
elements of the electron EDM operator (\ref{eq:eedm}) to the
scalar-pseudoscalar CP-odd operator (\ref{eq:r1-sp}) for eight different
atoms calculated using analytical formulas presented in the
appendix. The results for cesium and thallium are in good agreement
with the many-body calculations of Ref.~\cite{DF09}, the result for
Yb$^+$ is in excellent agreement with the many-body calculations of
Ref.~\cite{Quiney} for the YbF molecule. We stress that the Table
compares the ratios of the single-electron matrix elements obtained
with a simple analytical formula to the ratios of the EDMs obtained
with sophisticated many-body calculations.
These results provide an unambiguous link between the sign and
value of two different contributions to the EDM of atoms and molecules
which have a heavy atom from the Table~\ref{t:rcp}.

\subsection{EDM of polar molecules}

Polar molecules have strong inter-atomic electric field which enhance
the effect of electron EDM and lead to molecular EDM which are several
orders of magnitude larger than those in atomic systems. The
experimental search is in progress for YbF~\cite{HindsNature},
PbO~\cite{PbOe,PbO}, and ThO~\cite{ThOe} while other molecules are also
discussed in the literature (see, e.g.~\cite{Meyer08}). Interpretation of the
measurements requires molecular calculations. Table~\ref{t:mol} shows
the results of most recent or most accurate calculations for some polar
molecules. More detailed data are presented for YbF molecule for which
the EDM measurements were recently reported~\cite{HindsNature}.
The effects of electron EDM and scalar-pseudoscalar
CP-odd interaction are considered. The results are presented in terms
of the CP-odd parameters $W_d$ and $W_c$
\begin{eqnarray}
&&W_d = \langle \Psi_0|H_e|\Psi_0 \rangle/d_e, \\
&&W_c = \langle \Psi_0|H^{SP}|\Psi_0 \rangle/C^{SP}. 
\end{eqnarray}
To compare with other works one should keep in mind that most of them
present $W_S$ instead of $W_c$, where
\begin{equation}
W_S= \frac{2}{k_s}\langle \Psi_0|H^{SP}|\Psi_0 \rangle.
\label{eq:ws}
\end{equation} 
The constants $k_s$ and $C^{SP}$ of the strength of the CP-odd
interaction are related by 
$Zk_s=AC^{SP}$. Factor 2 in the definition of $W_S$ (\ref{eq:ws}) is
absent in some of the papers.

Most of the calculations of the molecular EDM include
only one of the CP-odd effects: that of the electron EDM or the
scalar-pseudoscalar interaction. We use the relations between the matrix
elements of the two CP-odd Hamiltonians to fill the gaps in the
table. Corresponding results are shown in bold. For example, 
according to Ref.~\cite{Meyer08} the effect of electron EDM in the YbF
molecule is $W_d=-15 \times 10^{24} \ {\rm Hz}/e \ {\rm cm}$. Using
the ratio $W_d/W_c = 114 \times 10^{18} /e \ {\rm
  cm}$ for Yb$^+$ from Table \ref{t:rcp} we found that the effect of
the scalar-pseudoscalar interaction is $W_c = -132$~kHz.

\begin{table}
\caption{$CP$-odd interaction constants $W_d$, and $W_c$ for some
  polar molecules and their ratios. Effective electric field $E_{\rm eff}$
  is presented together with $W_d$ ($\langle H_e \rangle = W_dd_e = -
  \mathbf{E_{eff}}\mathbf{d_e}$). 
The results of present paper are shown in bold.} 
\label{t:mol}
\begin{ruledtabular}
\begin{tabular}{lcccc}
 & $E_{\rm eff}$ & $W_d$ &  $W_c$ & $W_d/W_c$ \\ 
Molecule & (GV/cm) & ($10^{24} \ {\rm Hz}/e \ {\rm cm}$) & (kHz) & ($10^{18}/ e \ {\rm
  cm}$) \\
\hline
BaF        & 6.1\tablenotemark[1] & -1.5\tablenotemark[1] & {\bf -10} & {\bf 152} \\
           & {\bf 6.1} & {\bf -1.5}  & -10\tablenotemark[2] & {\bf 152} \\

YbF        & 31\tablenotemark[3] & -7.5\tablenotemark[3] & -59\tablenotemark[3] & 127 \\
           & 19\tablenotemark[4] & -4.6\tablenotemark[4] & -41\tablenotemark[4] & 112 \\
           & 26\tablenotemark[5] &-6.3\tablenotemark[5] & {\bf -55} & {\bf 114} \\
           & 26\tablenotemark[6] & -6.2\tablenotemark[6] & -54\tablenotemark[6] & 115 \\
           & 24\tablenotemark[7] & -5.8\tablenotemark[7] & -54\tablenotemark[7] & 108 \\
           & 25\tablenotemark[8] & -6.1\tablenotemark[8] & {\bf -53} & {\bf 114} \\

           & 32\tablenotemark[1] & -7.7\tablenotemark[1] & {\bf -68} & {\bf 114} \\
           & {\bf 21} & {\bf -5.2} & -46\tablenotemark[2] & {\bf 114} \\

HgF        & 95\tablenotemark[1] & -23\tablenotemark[1] & {\bf -226} & {\bf 90} \\

PbF        & -31\tablenotemark[1] & 7.5\tablenotemark[1] & {\bf 85} & {\bf 88} \\

PbO $a(1)$ & 25\tablenotemark[9] & -6.1\tablenotemark[9] & {\bf -69} & {\bf 88} \\ 
PbO $B(1)$ & 33\tablenotemark[9] & -8.0\tablenotemark[9] & {\bf -91} & {\bf 88} \\ 
PbO $a(1)$ & 23\tablenotemark[1] & -5.6\tablenotemark[1] & {\bf -64} & {\bf 88} \\ 

ThO        & 104\tablenotemark[1] & -25\tablenotemark[1] & {\bf -336} & {\bf 75}  \\
ThF$^+$    &  90\tablenotemark[1] & -22\tablenotemark[1] & {\bf -290} & {\bf 75}  \\
\end{tabular}
\tablenotetext[1]{Meyer and Bohn, Ref.~\cite{Meyer08}.}
\tablenotetext[2]{Nayak and Chaudhuri, Ref.~\cite{Nayak}.} 

\tablenotetext[3]{Kozlov and Ezhov, Ref.~\cite{KE94}.} 
\tablenotetext[4]{Titov {\em et al}, Ref.~\cite{TME96}.} 
\tablenotetext[5]{Kozlov, Ref.~\cite{K97}.} 
\tablenotetext[6]{Quiney {\em et al}, Ref.~\cite{Quiney}.} 
\tablenotetext[7]{Parpia, Ref.~\cite{Parpia}.} 
\tablenotetext[8]{Mosyagin {\em et al}, Ref.~\cite{MKT98}.} 

\tablenotetext[9]{Petrov {\em et al}, Ref.~\cite{Petrov}.} 
\end{ruledtabular}
\end{table}

\subsection{Extraction of the electron EDM and the parameter of the
  scalar-pseudoscalar CP-odd interaction from the EDM measurements for
  YbF and Tl}

Recent measurement of T and P violation in YbF molecule~\cite{HindsNature} 
combined with the Tl EDM measurement~\cite{Regan} and the data from
Table~\ref{t:rcp} allows to obtain independent limits on the electron EDM
and the CP-violating interaction. The values of the electron EDM extracted
from the experimental data for Tl~\cite{Regan} and
YbF~\cite{HindsNature} under assumption that there is no other
contribution to atomic/molecular EDM read
\begin{eqnarray}
{\rm Tl:}  & d_e=& (6.9 \pm 7.4) \times 10^{-28} \ e \ {\rm cm}, 
\label{eq:tl} \\
{\rm YbF:} & d_e=& (-2.4 \pm 5.7 \pm 1.5) \times 10^{-28} \ e \ {\rm
  cm}. \label{eq:ybf}
\end{eqnarray}
In fact, there are other contributions from the CP-odd
electron-nucleus interactions. Here we consider only the
scalar-pseudoscalar interaction (\ref{eq:s-p}). Other contributions
should be small due to the constrains obtained from the EDM measurements
for mercury~\cite{HgEDM}. Using data from Table~\ref{t:rcp} one can
rewrite (\ref{eq:tl},\ref{eq:ybf}) as
\begin{eqnarray}
{\rm Tl:}  & d_e + 1.2 \times 10^{-20} C^{SP}= \nonumber \\
& (6.9 \pm 7.4) \times 10^{-28} \ e \ {\rm cm},  \label{eq:tlc} \\
{\rm YbF:} & d_e + 8.8 \times 10^{-21} C^{SP}= \nonumber \\
& (-2.4 \pm 5.7 \pm 1.5) \times 10^{-28} \ e \ {\rm cm}. \label{eq:ybfc}
\end{eqnarray}
Solving these equations for $d_e$ and $C^{SP}$ leads to
\begin{eqnarray}
 &&d_e = (-2.8 \pm 3.0) \times 10^{-27} \ e \ {\rm cm}, \label{eq:de} \\
 &&C^{SP} = (3.0 \pm 3.0) \times 10^{-7} \label{eq:csp}.
\end{eqnarray}
The limit (\ref{eq:de}) for $d_e$ is significantly weaker than those
presented in (\ref{eq:tl}) and (\ref{eq:ybf}). This is because the
values of the ratios of the matrix elements of the two operators are
very close for Tl and Yb (see Table~\ref{t:rcp}). The most fortunate
case leading to strongest limits for $d_e$ and $C^{SP}$ would
correspond to very different values of the ratio, preferably with the
different sign. However, the formulas (\ref{eq:eedm}) and
(\ref{eq:r1-sp}) and the data in Table~\ref{t:rcp} show that this is
not possible. The ratio is always positive and slowly decrease with
$Z$. A slightly different value of the ratio should be expected for
the EDM of radium in the excited metastable $7s6d \ ^3D_2$ state.
The EDM of radium in this state is strongly
enhanced~\cite{DFG00} due to proximity of the $7s7p \ ^3P^o_1$ state
(the energy interval is $\sim 5 \ {\rm cm}^{-1}$) and is proportional
to the $\langle 7p_{3/2}|H_{CP}|6d_{3/2} \rangle$ matrix element of the
CP-odd interaction $H_{CP}$. However, the value of this matrix element
is strongly dominated by the core polarization effect which in turn is
mostly due to the $s_{1/2} - p_{1/2}$ matrix elements between core and
excited states. Therefore, the ratio of the two contributions is not
very different from what was considered above for Tl and YbF.

\subsection{EDM of Tl and Fr}

EDM of thallium due to electron EDM and scalar-pseudoscalar CP-odd
interaction was recently calculated in our paper~\cite{DF09}. The ratio
of the two contributions was found to be 83$d_e/(C^{\rm SP} 10^{-18} e
\ {\rm cm})$ which is in good agreement with the analytical result of
this paper: $W_d/W_c=89/( 10^{-18} e \ {\rm cm})$.
The EDM enhancement factor is $-582$~\cite{DF09} which is in very good
agreement with the value $-585$ found in earlier calculations by Liu and
Kelly~\cite{Liu}. The
EDM enhancement factor for thallium was also calculated in more recent
work of Ref.~\cite{Nataraj} and found to be -466 ($d_{\rm Tl}=-466d_e$) which
is about 25\% smaller than the results mentioned above.
The EDM of thallium due to the
scalar-pseudoscalar CP-odd interaction calculated by the same group
earlier ($d_{\rm Tl} = -4.06 \times 10^{-18} C^{\rm SP} e \ {\rm
  cm}$~\cite{Sahoo08}) is also smaller than in other calculations of
Ref.~\cite{DF09} and \cite{Pendrill}. On the other hand, the ratio of
the two contributions 
is 115 (in units of $d_e/(C^{SP} 10^{-18} \ e \ {\rm cm})$)
which is significantly larger than the analytical and numerical
values of 89 and 83 (same units, see Table~\ref{t:rcp}). 

EDM enhancement factor for francium was calculated in Ref.~\cite{BDFM}
and found to be 910(46): $d({\rm Fr}) = 910(46) d_e$. This value is in
good agreement with the value of 894.93 found in more recent
calculations~\cite{Mukherjee}. Using the value $d({\rm Fr}) = 910(46)
d_e$ and  the ratio of the 
electron-EDM matrix element to the scalar-pseudoscalar interaction
presented in Table~\ref{t:rcp} we can now reconstruct
another contribution to the EDM of francium:
\begin{equation}
d({\rm Fr}) = 11\times 10^{-18}C^{\rm SP} e \ {\rm cm}.
\label{eq:frcsp}
\end{equation}
It is interesting to note that the calculation of the nuclear
spin-dependent PNC amplitude between hyperfine components of the
ground state of francium presented in Ref.~\cite{Porsev01} involves
the same sum as the expression (\ref{eq:edm}) for the EDM of the
atom. Therefore, we can reconstruct the EDM enhancement factor for Fr
using the results of Ref.~\cite{Porsev01} and proportionality between
matrix elements of the nuclear spin-dependent interaction
(\ref{eq:sdpnc-sp}) and the electron EDM operator (\ref{eq:eedm}).
This leads to $d({\rm Fr}) = 854 d_e$ which is in good agreement
with the value $d({\rm Fr}) = 910(46) d_e$ of Ref.~\cite{BDFM}. This
is a good consistency test of both calculations.

Reading the papers citing our works (see,
e.g. \cite{Nataraj,Mukherjee}) reveals the need to clear some points
about our method of calculations.
For example, the atomic
electric field interacting with electron EDMs is 
calculated in \cite{DF09} and \cite{BDFM} as a derivative of the total
potential which includes both nuclear and electron parts. The formula
$\mathbf{E} = Ze\mathbf{r}/r^3$ for the leading contribution to the atomic
electric field presented on first page of~\cite{BDFM} may make an impression
that only nuclear field is included. However, the formula (2) few
lines below clearly includes screening functions $Q(r)$ and $P(r)$ for
both the nuclear Coulomb and the external electric field. 
By the way, the inclusion of the electron electric field change the
matrix elements of the electron EDM for thallium by 0.4\% only. This
is because main contribution comes from short distances where the electron
electric field is small since the electron potential rapidly tends to
a constant inside the $1s$ orbital.
Therefore, we include only nuclear electric field in the analytical
analysis (formula (\ref{eq:eedm}) in the appendix) while keeping both
contributions in the numerical calculations.

\begin{figure}
\epsfig{figure=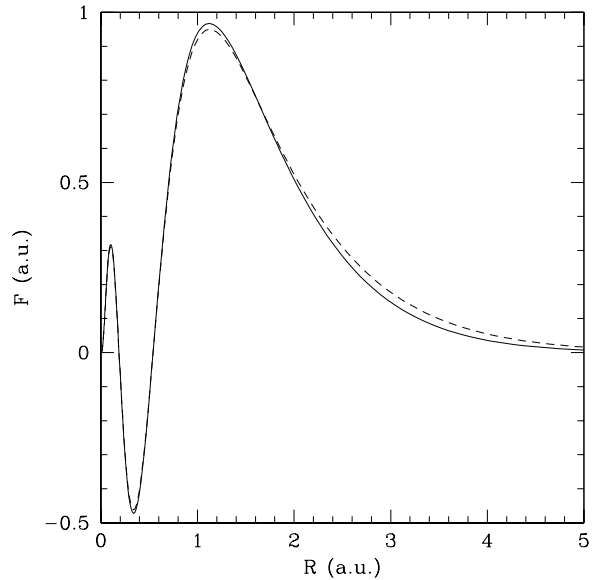,scale=0.40}
\caption{The $5d_{5/2}$ core function of Tl calculated in $V^{N-3}$
  (solid line) and $V^{N-1}$ (dashed line) approximations.}
\label{fig:ff}
\end{figure}

The authors of~\cite{Nataraj} claim that atomic core is strongly
contracted in the $V^{N-3}$ starting approximation used in our
calculations~\cite{DF09}.  In fact, it is not. 
Fig.\ref{fig:ff} shows the outermost $5d_{5/2}$ core function 
of Tl calculated in the $V^{N-3}$ and $V^{N-1}$ approximations. The
difference between the functions is very small. 
This is due to the fact that the valence $6s$ and $6p$ electrons 
are located outside of atomic core. Their charge
distribution creates almost constant potential and no electric field
inside the sphere where all inner electrons are located. Therefore,
the valence electrons have practically no effect on the core wave functions
(see~\cite{VNM} for a detailed discussion).  
The change is even smaller for other core functions. 
The core functions enter the
configuration interaction (CI) Hamiltonian  via core potential
$V_{core}$ to which all core electrons contribute (from $1s$ to
$5d$). The difference for 
$V_{core}$ in the $V^{N-3}$ and $V^{N-1}$ approximations is very small~\cite{VNM}.
Moreover, the corresponding corrections to the configuration
interaction (CI) Hamiltonian have been included in~\cite{DF09}
using the many-body perturbation theory approach. 

As it is well known, the eigenstates of a Hamiltonian do not depend on
the basis one uses.
The valence states are indeed different in the $V^{N-3}$ and $V^{N-1}$
approximations. However, this should have no effect on the final
results as long as the saturation of the basis for valence states is
achieved. There are 
only two conditions the basis states must satisfy: (a) they must be
orthogonal to the core, and (b) they must constitute a complete set
of states. Therefore, it does not matter whether valence states are
calculated in the $V^{N-3}$ or $V^{N-1}$ potential or by any other method
(e.g., a popular basis $\psi_n(r)=r^n\psi_0(r)$~\cite{KPF96}), the final
results should be the same. If there is any difference in the results, the
most likely reason for this is the incompleteness of the basis set. 

In spite of no difference in final results there is a good reason
for the use of the $V^{N-3}$ approximation -- the simplicity and good
convergence of the many-body perturbation theory (MBPT) for the
core-valence correlations. This is the only approximation which has no
{\em subtraction diagrams}~\cite{DFK96}. To be more precise, we should
state that the condition for the absence of the subtraction diagrams
is that the potential used to calculate single-electron core states
and the potential of the core in the CI Hamiltonian are exactly the
same. In the case of thallium atom treated as a
three-valence-electrons system this corresponds to the $V^{N-3}$
approximation. When any other approximation is used one has to include
the subtraction diagrams. 

Large energy denominators 
suppress the value of the correlation terms in the $V^{N-3}$
approximation ensuring good convergence of the MBPT~\cite{VNM}.
There must be large cancelation between subtraction and other
diagrams to ensure the same final results if any other initial
approximation is used. This is very similar to the well known fact that
the Hartree-Fock basis is the best choice for any MBPT
calculations. Zero-order results might be better in some other
approximation, however, strong cancelation between subtraction and
other diagrams would lead to poor convergence of the MBPT.

The authors of~\cite{Nataraj} claim that the major drawback of our
work~\cite{DF09} is the inclusion of the core-valence correlations
in the second order only. However, the correlations between the valence
electrons and core electrons below the $6s$ state are small which is evident
from the fact that their inclusion change the EDM of Tl by 3\%
only~\cite{DF09}.
Therefore, only the 
correlations between three valence electrons should be treated
to all orders. This is done in~\cite{DF09} to a very high precision
using the CI technique.

\section{Conclusion}

We have demonstrated the proportionality relations between the nuclear-spin
dependent and spin-independent PNC effects in atoms and the
scalar-pseudoscalar and the electron EDM contributions to the EDM of
atoms and molecules. The relations are given by the simple analytical
formulas and can be used to express one symmetry breaking effect
through another. Using these relations and accurate calculations of
the spin-independent PNC we have confirmed earlier interpretations of
the nuclear anapole measurements in cesium and thallium. We have also
confirmed the ratio of the scalar-pseudoscalar CP-odd and electron EDM
contributions to the EDM of Cs, Tl and other atoms and some polar
molecules. Using the relations we found 
the scalar-pseudoscalar contribution to the EDM of francium atom and
the scalar-pseudoscalar and electron EDM contributions to the EDM of
many polar molecules. Using experimental limits on EDMs of thallium
and YbF we found model-independent limits on the electron EDM and the
constant of the scalar-pseudoscalar CP-odd interaction. 

\subsection*{Acknowledgments.}

The authors are grateful to E. A. Hinds for stimulating discussions.
The work was supported in part by the Australian Research Council.

\appendix
\section{Matrix elements of the weak interaction}
\subsection{Wave function}
We use single-electron wave functions in a form
\begin{equation}
    \psi(r)_{njlm}=\frac{1}{r}\left(\begin {array}{c}
    f_{v}(r)\Omega(\mathbf{n})_{\mathit{jlm}}  \\[0.2ex]
    i\alpha g_{v}(r)  \widetilde{ \Omega}(\mathbf{n})_{\mathit{jlm}}
    \end{array} \right),
\label{psi}
\end{equation}
where $n$ is the principal quantum number and an index $v$
replaces the three-number set $n,j,l$; $\alpha$ is the fine structure
constant. 

We need $s_{1/2}$ and $p_{1/2}$ wave functions inside the nucleus to
calculate matrix elements. Following Ref.~\cite{FG02} we assume
uniform distribution of electric charge inside a sphere of radius
$R_N$. Taking formulas for the $s_{1/2}$ and $p_{1/2}$ wave functions
  from Ref.~\cite{FG02} and keeping terms up to $Z^2\alpha^2$
we come to the following expressions
\begin{eqnarray}
&& f_s(x) = A_s x\left(1-\frac{3}{8}Z^2\alpha^2x^2\right), \nonumber \\
&& g_s(x) = -A_s Z\frac{x^2}{2}\left(1-\frac{x^2}{5}\right),  \label{eq:sp12} \\
&& f_{p_{1/2}}(x) =
A_p Z\alpha^2\frac{x^2}{2}\left(1-\frac{x^2}{5}\right), \nonumber \\
&& g_{p_{1/2}}(x) = A_p x\left(1-\frac{3}{8}Z^2\alpha^2x^2\right). \nonumber 
\end{eqnarray}
Here $x=r/R_N$, $R_N$ is nuclear radius.
 $A_s$ and $A_p$ are normalization factors. They  can be found by
matching (\ref{eq:sp12}) to atomic wave functions at short distances
outside of the nucleus (see, e.g.~\cite{Khrip}). 
The use of the semiclassical wave functions presented in~\cite{Khrip}
leads to approximate expressions
\begin{eqnarray}
&& A_s = \frac{(1+\gamma)(2ZR_N)^{\gamma}}{(1 -
  0.375Z^2\alpha^2)\sqrt{Z}\Gamma(2\gamma + 1)\nu_s^{1.5}} \nonumber \\
&& A_p = \frac{\sqrt{Z}(2ZR_N)^{\gamma}}{(1 -
  0.375Z^2\alpha^2)\Gamma(2\gamma + 1)\nu_p^{1.5}} \label{eq:ab},
\end{eqnarray}
where $\gamma = \sqrt{1-Z^2\alpha^2}$, $\Gamma$ is gamma-function,
$\nu$ is effective principal quantum number: single-electron energy of
state $n$ is $\epsilon_n = -1/(2\nu_n^2)$.
Note that normalization factors (\ref{eq:ab}) are needed only to
compare the matrix elements of the electron EDM to other matrix
elements. For cases not involving electron EDM there is exact
cancellation of the normalization factors. Therefore, their
uncertainty does not contribute to the uncertainty of final results.

\subsection{PNC matrix elements}

Matrix element of the SI PNC $H_{\rm SI}$ interaction (first term in
(\ref{e1})) is 
\begin{eqnarray}
&&\langle \kappa_1 |H_{\rm SI}| \kappa_2 \rangle =
i\frac{G_FQ_W}{2\sqrt{2}} \times \label{eq:sipnc} \\
&& \alpha \delta_{-\kappa_1,\kappa_2} 
\int (f_1g_2-g_1f_2)\rho(r)dr. \nonumber
\end{eqnarray}
Substituting (\ref{eq:sp12}) for the $s_{1/2}-p_{1/2}$ transition and
assuming uniform nuclear charge distribution we get
\begin{eqnarray}
&&\langle s_{1/2} |H_{\rm SI}| p_{1/2} \rangle =
i\frac{G_FQ_W}{2\sqrt{2}}  \times \label{eq:sipns-sp} \\
&&\frac{\alpha \rho_0 A_sA_p R_N}{3}\left(1 -
  0.34Z^2\alpha^2\right). \nonumber
\end{eqnarray}
Here $A_s$ and $A_p$ are given by (\ref{eq:ab}) and 
\begin{equation}
  \rho_0 = \frac{3}{4\pi R_N^3}.
\label{eq:rho0}
\end{equation}
Matrix element of the SD PNC ${\bm b}$ operator (\ref{eq:hsdp}) is
\begin{eqnarray}
&&\langle \kappa_1 ||{\bm b}|| \kappa_2 \rangle =
-i\frac{G_F\varkappa}{\sqrt{2}}  \alpha \langle
-\kappa_1||C_1||\kappa_2 \rangle \times \label{eq:sdpnc} \\
&& \int\left((\kappa_2-\kappa_1+1)f_1g_2 -
  (\kappa_1-\kappa_2+1)g_1f_2\right) \rho(r)dr. \nonumber 
\end{eqnarray}
The reduced matrix element of the spherical harmonic $C_k$ is
\begin{eqnarray}
&\langle \kappa_a||C_k||\kappa_b \rangle = 
(-1)^{j_b+1/2}\sqrt{(2j_a+1)(2j_b+1)} \nonumber \\
& \times \xi(l_a+l_b+k)\left( \begin{array}{ccc} j_b & j_a & k \\
-1/2 & 1/2 & 0 \end{array} \right) . \label{eq:ck} \\
& \xi(x)= \left\{ \begin{array}{ccccc} 1, & {\rm if} &x& {\rm is}&{\rm even} \\
0, & {\rm if} &x& {\rm is}&{\rm odd} \end{array} \right. \nonumber
\end{eqnarray} 
For the $s_{1/2}-p_{1/2}$ transition and uniform nuclear charge distribution
we have 
\begin{eqnarray}
&&\langle s_{1/2} ||{\bm b}|| p_{1/2} \rangle =
-i\frac{G_F\varkappa}{\sqrt{2}}  \times \label{eq:sdpnc-sp} \\
&&\alpha \sqrt{\frac{2}{3}}\rho_0 A_sA_p R_N\left(1 -
  0.41Z^2\alpha^2\right). \nonumber
\end{eqnarray}
The ratio of (\ref{eq:sdpnc-sp}) to (\ref{eq:sipns-sp}) is
\begin{eqnarray}
&&\langle s_{1/2} ||{\bm b}|| p_{1/2} \rangle/
\langle s_{1/2} |H_{\rm SI}| p_{1/2} \rangle \approx \label{eq:r1} \\
&& 6\sqrt{\frac{2}{3}}\frac{\varkappa}{(-Q_W)}\left(1-0.073Z^2\alpha^2\right).
\nonumber
\end{eqnarray}

\subsection{Scalar-pseudoscalar CP-odd interaction}  

The Hamiltonian of the scalar-pseudoscalar electron-nucleon (T,P)-odd
interaction can be written as \cite{Ginges}
\begin{equation}
  H^{\rm SP} = i\frac{G_F}{\sqrt{2}}AC^{\rm SP}\gamma_0\gamma_5\rho_N(r),
\label{eq:s-p}
\end{equation}
where $G_F$ is the Fermi constant, $A=Z+N$ is the nuclear mass number, 
$Z$ is the number of protons and $N$ is the number of neutrons.
$C^{\rm SP}=(ZC^{\rm SP}_p +NC^{\rm SP}_n)/A$, where $C^{\rm SP}_p$
and  $C^{\rm SP}_n$
are the parameters of the scalar-pseudoscalar (T,P)-odd interaction
for protons and neutrons,
$\gamma_n$ are the Dirac matrices.
\begin{eqnarray}
  \langle a|H^{\rm SP}|b \rangle &=& - \frac{G_F}{\sqrt{2}}\alpha
  AC^{\rm SP} \nonumber \\
&&\times \delta_{-\kappa_a,\kappa_b}
\int(f_ag_b+g_af_b)\rho_N dr. 
\label{eq:r-sp}
\end{eqnarray}
Substitution of the functions (\ref{eq:sp12}) leads to
\begin{eqnarray}
 && \langle s_{1/2}|H^{\rm SP}|p_{1/2} \rangle = \label{eq:r1-sp} \\
&& - \frac{G_FAC^{\rm SP}}{3\sqrt{2}}\alpha A_sA_p\rho_0R_N \left(1
    -0.56Z^2\alpha^2\right).
\nonumber
\end{eqnarray}
Here $A_s$ and $A_p$ are given by (\ref{eq:ab}) and $\rho_0$ is given
by (\ref{eq:rho0}).

\subsection{Electron EDM}

The Hamiltonian for the electron EDM interacting with internal atomic
electric field $\mathbf{E}_{\rm int}$ can be written
as~\cite{Khrip}
\begin{equation}
  H_e = -d_e(\gamma_0-1)\mathbf{\Sigma}\cdot\mathbf{E}_{\rm int},
\label{eq:he2}
\end{equation}
where \begin{equation}\mathbf{\Sigma} = \left(\begin{array}{cc}
      \mathbf{\sigma} & 0 
  \\ 0 & \mathbf{\sigma} \end{array}\right),  \nonumber \end{equation} and
$\mathbf{E}_{\rm int}$ is internal atomic electric field.
The $s_{1/2}-p_{1/2}$ matrix element can be found in Ref.~\cite{F76}:
\begin{equation}
\langle s_{1/2} |H_e| p_{1/2} \rangle =
 -\frac{4Z^3\alpha^2d_e}{\gamma(4\gamma^2-1)(\nu_s\nu_p)^{3/2}}.
 \label{eq:eedm} 
\end{equation}

\end{document}